\documentclass[pdflatex,sn-basic]{sn-jnl}% Basic Springer Nature Reference Style/Chemistry Reference Style

\usepackage{algorithm}%
\usepackage{algorithmicx}%
\usepackage{algpseudocode}%
\usepackage{amsmath,amssymb,amsfonts}%
\usepackage{amsthm}%
\usepackage[title]{appendix}%
\usepackage{bm}
\usepackage{booktabs}%
\usepackage{graphicx}%
\usepackage{listings}%
\usepackage{manyfoot}%
\usepackage{mathrsfs}%
\usepackage{multirow}%
\usepackage{textcomp}%
\usepackage{xcolor}%
\theoremstyle{thmstyleone}%
\theoremstyle{thmstyletwo}%

\theoremstyle{thmstylethree}%

\begin{document}

\title[DTB for TDoA location systems]{Handling systematic node biases in TDoA positioning systems}

\author*[1,2]{\fnm{Miquel} \sur{Garcia-Fernandez}}\email{miquel.garcia.fernandez@upc.edu}

\affil*[1]{\orgname{UPC/IonSat}, \orgaddress{\city{Barcelona}, \postcode{08034}, \state{Catalunya}, \country{Spain}}}
\affil*[2]{\orgname{MGFernan consulting}, \orgaddress{\city{Sant Andreu de la Barca}, \postcode{08740}, \state{Catalunya}, \country{Spain}}}


\abstract{
Node-specific hardware biases represent significant error source in wireless
Time of Arrival (ToA) and Time Difference of Arrival (TDoA) positioning.
While bias estimation and treatment are well-established in Global Navigation
Satellite Systems (GNSS), their application in wireless Positioning,
Navigation, and Timing (PNT) remains a significant
challenge. This work proposes a methodology, based on GNSS
data processing, to mitigate these \emph{systematic biases}. Specifically, the concept of
Differential Transmitter Bias (DTB) is introduced to account for node-specific
hardware variations. These $DTB$ are in fact analogous to the Differential Code
Bias (DCB) used in GNSS. This paper includes an analysis based on real data
from a network of wireless nodes employed in an indoor navigation context and
shows that the $DTB$ exhibits remarkable stability, allowing it to be treated as
a constant during an entire positioning session, with small variations as low as 2m,
which is in the same uncertainty level as the noise of the measurements used by
a TDoA-based positioning engine. In an operating positioning service context, these $DTB$ would
be transmitted to the user, similarly to GNSS, where satellite
clock corrections are provided through navigation messages. This would allow the user to
calibrate the TDoA measurements in order to obtain positioning with errors on
the meter level.
}


\maketitle

\section{Introduction}\label{sec::introduction}

Hardware and clock biases are fundamental to Global Navigation Satellite System
(GNSS) measurements, and well-established methodologies exist for their treatment
(e.g., \cite{haakansson2017review}).  However, in wireless and alternative
Positioning, Navigation, and Timing (PNT) such as those based on 5G, Ultrawide-Band and Wi-Fi,
this topic has received considerably less attention. Most of the works
that study biases in those systems focus on biases originating from Non-Line-of-Sight
(NLOS) (e.g., \cite{torsoli2022selection}, \cite{christopher2021characterizing}, \cite{klus2024robust}, \cite{marano2010nlos},
\cite{prieto2009nlos}) and receiver (i.e. user equipment)
clock synchronization. In the case of clock synchronization, some works based on
simulations assume perfect synchronization (e.g., \cite{yesilyurt20235g}),
effectively nullifying clock errors. However, real-world receiver clocks, particularly
cost-effective ones, are subject to drift and jitter. Another type of biases
are the ones introduced in the estimated state computed within filters
like the Extended Kalman Filter (EKF), but they are independent from the underlying
measurement accuracy (e.g., see \cite{sriram2016tdoa} for Time Difference of Arrival (TDoA) positioning).

In the case of Wi-Fi Round Trip Time (RTT) technology, similar hardware biases are also present
(e.g., \cite{rana2024enhanced}, \cite{gentner2020wifi}). These biases were also
analyzed in \cite{garcia2021accuracy} in the context of access point (node) position
estimation using RTT, while Horn et al. (\cite{horn2024round}) addressed them in
a rover positioning scenario through measurement calibration.

Regarding the receiver clock synchronization, TDoA can be used to completely remove
its contribution from the measurements. TDoA-based positioning is in fact analogous
to the \emph{single difference} formulation in GNSS and this simplifies the positioning
engine. However, TDoA does not remove any potential node biases present in the measurements.

Although NLOS-induced biases are widely studied in terrestrial wireless
positioning, the literature on \emph{systematic error} due to hardware biases
originating in the nodes in wireless ToA/TDoA is sparse. \cite{cui2016direction}
identified a \emph{systematic error} beyond NLOS and proposed
estimating this bias alongside position, similar on how it is done in GNSS.
However, they also acknowledged the difficulty in calibrating inter-node biases,
the focus of the present paper. Other works (\cite{zhao2021learning}), relied
on the usage of machine learning approaches, like those using Neural Networks,
to estimate these node \emph{systematic biases}.

In this context, this paper presents a methodology for treating these systematic
node biases, drawing upon established GNSS techniques. Specifically, this work
proposes combining TDoA, to eliminate receiver clock errors, with the concept of
\emph{Differential Transmitter Bias} (or $DTB$, for short), based on the concept
of \emph{Differential Code Bias} ($DCB$) used in GNSS.

To illustrate the concepts and methodology used in this paper, a publicly available
data extracted from the Indoor Position and Indoor Navigation (IPIN) conference
competition has been used \cite{anagnostopoulos2024evaluating}.
This data has been used to characterize and model the
$DTB$ as well as illustrate its usage in a positioning engine.

\section{Methodology}

\subsection{Measurement model and DTB}\label{meas_model_and_dtb}

Positioning systems, such as Global Navigation Satellite Systems (GNSS), that
rely on Range, or Time of Arrival (ToA), measurements, utilize trilateration
to determine the position of a receiver (rover) based on \emph{pseudo}-ranges.
These \emph{pseudo}-ranges, however, contain more than just the geometric distance
between the rover and the nodes (satellite in the case of GNSS). They also include
several undesirable terms. In the GNSS context, these include receiver and satellite
clock biases, along with other factors such as atmospheric effects. In alternative
wireless Positioning, Navigation, and Timing (PNT) systems like 5G or Ultra-Wideband (UWB),
receiver (rover) and transmitter (node) biases are also present. While these biases
tend to exhibit slow variations (remaining nearly constant within short time intervals),
they must be considered somehow within the positioning engine. In this work, the
\emph{pseudo}-ranges ($P$) between a rover ($r$) and a node ($n$) in a wireless
terrestrial positioning system will be modelled as:

\begin{equation}
P_r^n = \rho_r^n + b_r - b^n + \varepsilon_P
\label{eq::toa}
\end{equation}

where:

\begin{itemize}
\item $\rho_r^n$ is the geometric Euclidean distance between the rover $r$ and node $n$
    positions

\begin{equation}
\rho = \sqrt{(x_r-x^n)^2+(y_r-y^n)^2+(z_r-z^n)^2}
\label{eq::rho}
\end{equation}

\item $b_r$ and $b^n$ are the hardware (or clock) biases of the rover and node
    respectively. The sign convention of the biases are taken from GNSS convention,
    where the receiver biases are added and the satellite ones are substracted.
\item $\varepsilon_P$ includes the thermal noise and other unmodelled terms (such as e.g. multipath).
\end{itemize}

\begin{figure}[h]
\centering
\includegraphics[width=0.9\textwidth]{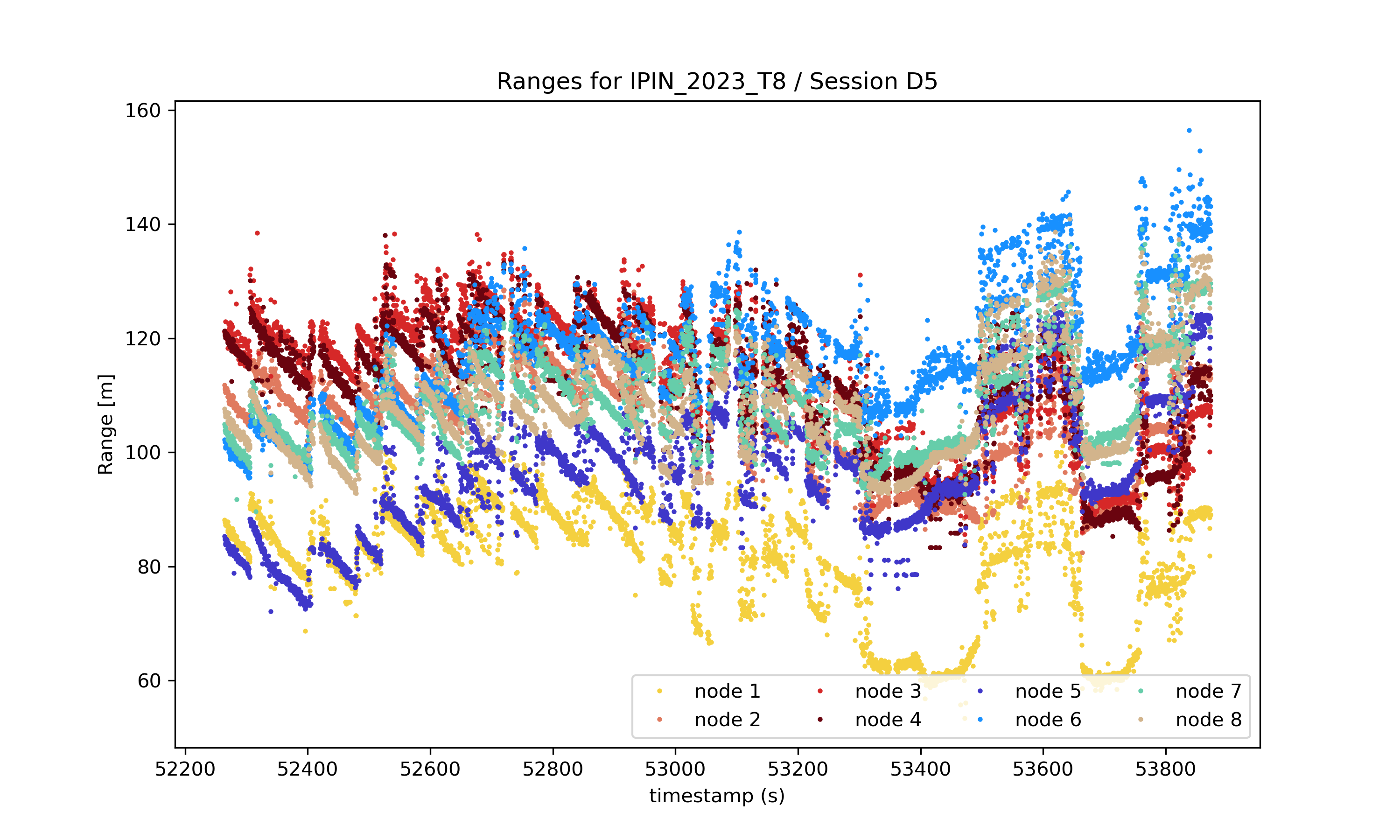}
\caption{Time of Arrival (ToA) measurements converted to distance (ranges) recorded
by the rover for all the nodes in a sample session}\label{fig::toa}
\end{figure}

\begin{figure}[h]
\centering
\includegraphics[width=0.9\textwidth]{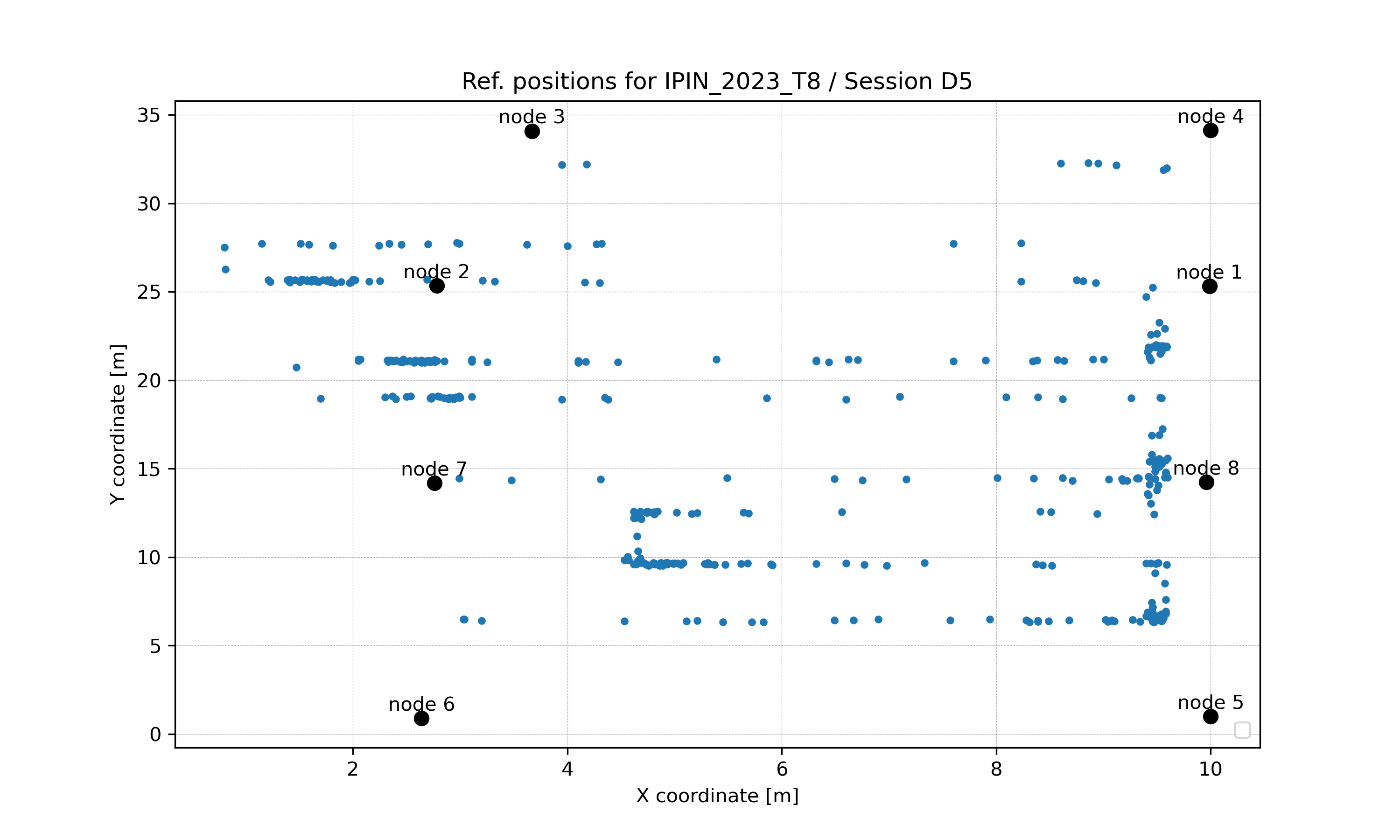}
\caption{Reference positions within the track followed by the rover during the Session D5
of the IPIN 2023 competition dataset. The position of the nodes are also shown in
the figure as black dots}\label{track_fig}
\end{figure}

As an example of these measurements, Figure \ref{fig::toa} includes the \emph{pseudo}-ranges
recorded by a moving rover for all the nodes during the IPIN 2023 campaign (data session \texttt{D5}).
From the range plots, one evident feature that can be seen are the jumps (\emph{sawtooth}-like shape)
that do not seem to correspond with actual movement (see reference points along the the track in Figure \ref{track_fig}).
Moreover, these jumps occur for all nodes, which indicates that they are due to the rover receiver
(presumably a receiver clock/oscillator issue). Finally, range values are much larger
($> 60m$) than the node relative distances ($\simeq 20 m$), which also points out to a bias issue.

In positioning algorithms such as GNSS, the receiver clock can be estimated along
with the three position coordinates, provided that enough satellites are in view (i.e. at least 4,
see for instance \cite{misra2011global}). An alternative is to entirely remove the
receiver clock by forming the so called \emph{single-differences} (noted with the
symbol $\nabla$), in which a reference node ($m$) is selected to refer all the other
measurements to it. In practice, processing
\emph{single-differences} observables is equivalent to state
that the Time Difference of Arrival (TDoA) measurements are processed. Therefore,
the terms \emph{single-differences} and TDoA will be used indistinctibly in this paper.
Using the pseudo-range definition in Eq. \ref{eq::toa}, the \emph{single-differenced}
range (or TDoA) between rover $r$ and node $n$ considering the reference node $m$
($\nabla P_r^{n,m}$), can be expressed as:

\begin{eqnarray}
\nabla P_r^{n,m} & = & P_r^n - P_r^m \nonumber \\
& = & \rho_r^n - \rho_r^m - b^n + b^m \nonumber \\
& = & \nabla \rho_r^{n,m} - b^n + b^m
\label{eq::tdoa}
\end{eqnarray}

\begin{figure}[h]
\centering
\includegraphics[width=0.9\textwidth]{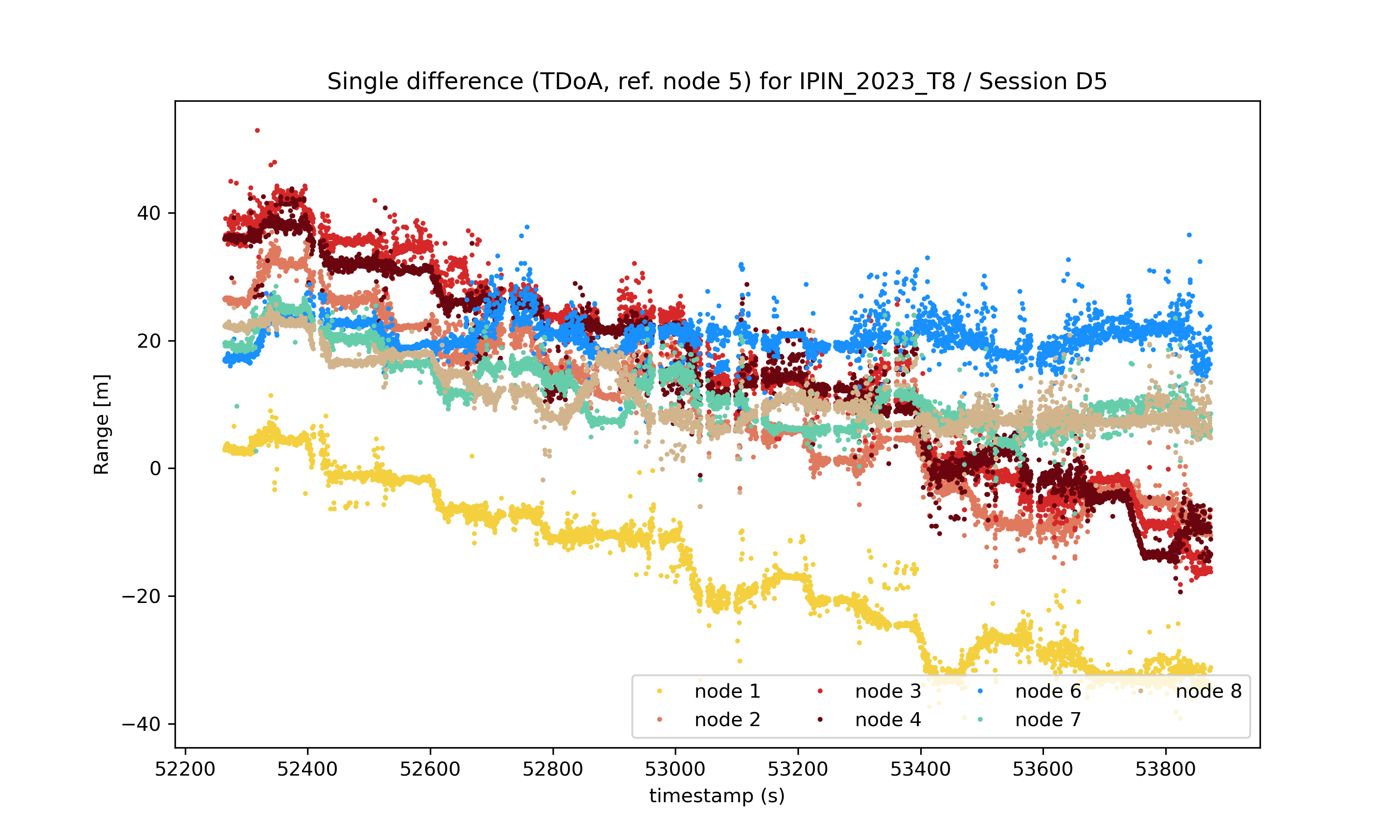}
\caption{Single difference/TDoA measurements of Session D5 of the IPIN competition dataset, built
as the difference of TOA measurements relative to a reference node}\label{fig::tdoa}
\end{figure}

Notice how, by forming \emph{single-differences}, the common terms of the rover (i.e.
clock/hardware bias $b_r$) vanish. Figure \ref{fig::tdoa} shows an example of
the single-difference/TDoA measurements for the same dataset shown in Figure \ref{fig::toa}.
Notice how both the excursions of values as well as the mean values are substantially
smaller than the corresponding TOA measurements. The remaining signal present in those
observables is due to the contribution of the single-differenced geometry
($\nabla \rho_r^{n,m}$).

Since the dataset contains the reference positions of the rover (see Figure \ref{track_fig}),
it is possible to compute the actual geometry (i.e. $\nabla \rho_r^{n,m}$),
which in turn allows estimating the \emph{Differential Transmitter Biases}
between the node $n$ and the referencenode $m$ (or $DTB_{n,m}$):

\begin{eqnarray}
DTB_{n,m} & = & \nabla P_r^{n,m} - \nabla \rho_r^{n,m}  \nonumber \\
& = & - b^n + b^m
\label{eq::dtb}
\end{eqnarray}

\begin{figure}[h]
\centering
\includegraphics[width=0.9\textwidth]{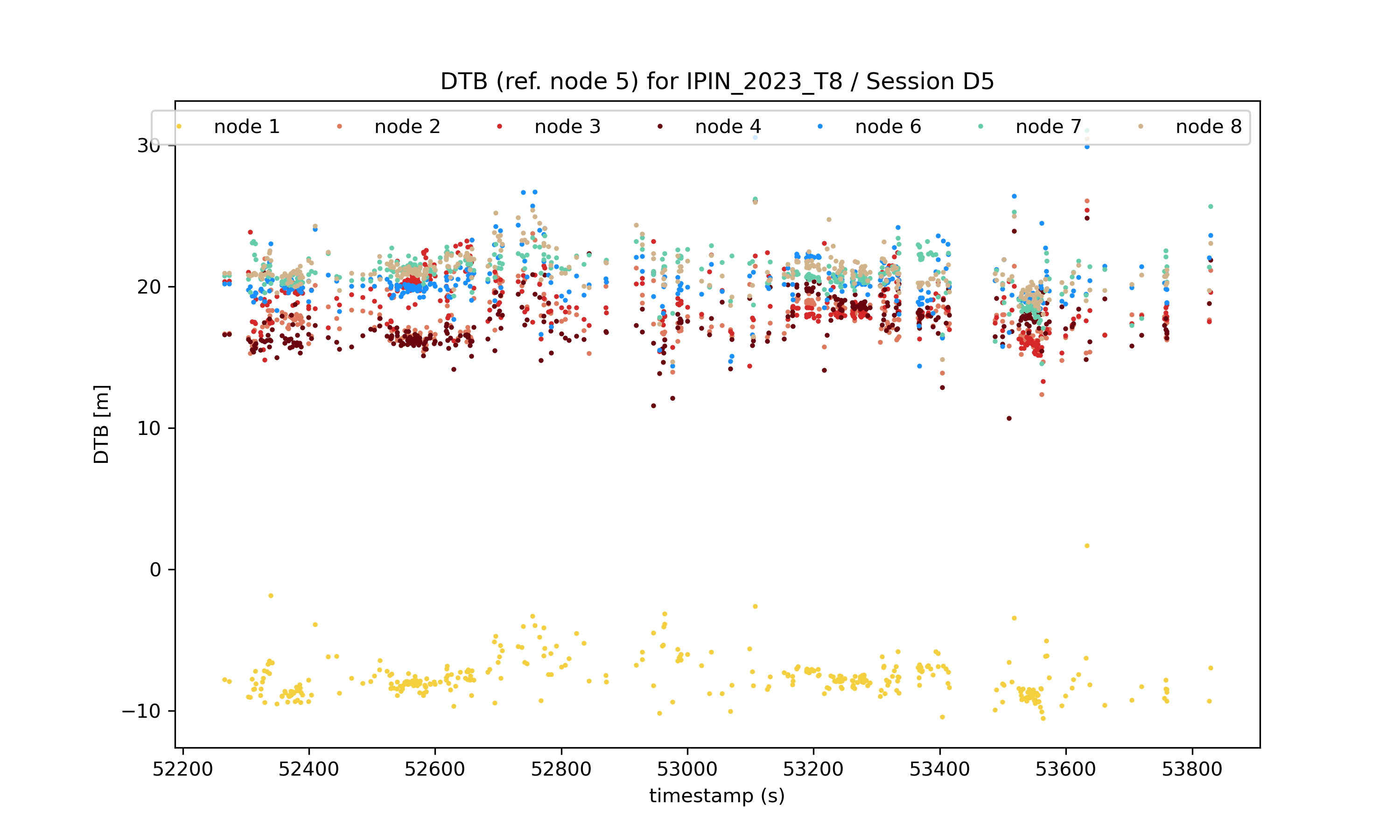}
\caption{$DTB_{n,m}$ of Session D5 of the IPIN 2023 competition dataset, built
as the difference of TDoA measurements and the actual geometry.}
\label{fig::dtb}
\end{figure}

An example of the instantaneous $DTB_{n,m}$ for the TDoA measurements is shown in
Figure \ref{fig::dtb}. Note that the time series indicate that the DTB can be
modelled as a constant bias term by computing the average. While some authors
attribute measurement biases primarily to non-line-of-sight (NLOS) observations
(\cite{torsoli2022selection}), the constant DTB observed in Figure \ref{fig::dtb}
suggests that, in this particular dataset, hardware-related biases may be the dominant factor.

The averaged $DTB_{n,m}$ can be subsequently used in a positioning algorithm
to calibrate the TDoA measurements, thus avoiding the need to estimate any
bias altogether. In an operational system, the $DTB_{n,m}$ would have to be
computed by the service provider and then disseminated to the user, in a similar
manner as the GNSS control segments do it for their final users (i.e. satellite
clocks and time group delays are broadcasted in the signal in space via the
navigation message).

Now that the methodology has been introduced for a single dataset, the Results section
(Section \ref{sec::results}) shows the statistics for all nodes and all datasets.

\subsection{Positioning Engine}\label{sec::posengine}

In order to compute the rover position using the TDoA measurements, an Extended
Kalman Filter (EKF) will be used to handle the non-linearity of the problem derived
from the geometric range equation. The formulation of the EKF for the specific case
of navigation and positioning has been widely used in previous works and thus
will not be fully reproduced here (the reader is referred to \cite{misra2011global}),
only the following key points to configure it will be mentioned so that the reader can reproduce
the results in other EKF implementations, such as the one included in \cite{mgf2025pygnss}:

\begin{itemize}
\item {\bf Apriori}, the position apriori will be computed as the average position of
the nodes and its uncertainty defined by the dispersion of the average (i.e. standard deviation).
\item The observation model consists in two parts:
\begin{itemize}
\item {\bf measurement model}, that transfers the state (i.e. two dimensional position $(x_r, y_r)$)
to model observations (i.e. $\nabla \tilde P_r^{n,m}$). This model is based on
Equation \ref{eq::tdoa}, where the biases are replaced by the $DTB$ (that need to be
provided to the engine along with the measurements):

\begin{eqnarray}
\nabla \tilde P_r^{n,m} = \nabla \rho_r^{n,m} + DTB_{n, m}
\label{eq::tdoa_model}
\end{eqnarray}

\item {\bf Jacobian matrix} with the partials ($\bm{H}$). This matrix is in fact equivalent
to the Jacobian matrix for a ToA (or range) based positioning system such as GNSS
(see equation 6.8 of \cite{misra2011global}), and can be formulated as:

\begin{equation}
\bm{H} =
\begin{pmatrix}
    \frac{x^1-x_r}{\rho_r^1} & \frac{y^1-y_r}{\rho_r^1} \\
    \frac{x^2-x_r}{\rho_r^2} & \frac{y^2-y_r}{\rho_r^2} \\
    \vdots & \vdots \\
    \frac{x^N-x_r}{\rho_r^N} & \frac{y^N-y_r}{\rho_r^N}
\end{pmatrix}
\label{eq::jacobian}
\end{equation}

where $N$ is the total number of nodes (and hence the total number of measurements
in each epoch if all nodes are visible), $(x^i, y^i)$ are the coordinates of the
$i$-th node and $\rho_r^i$ is the geometric range between the rover and the $i$-th
node. As a side node, in a ToA positioning system where the biases are not cancelled,
a column of $1$s should be required for the bias estimation and additional columns
would be required to estimate the node biases. However, estimating the biases
usually leads to ill-posed (rank deficient) linear systems, thus justifying the adoption
of differenced algorithms such as the one proposed in this work.

\end{itemize}
\item {\bf State propagation} will be based on a simple stochastic propagation model
driven by a $\Phi$ matrix, which can be constructed from the identity matrix:

\begin{equation}
\bm{\Phi} =
\begin{pmatrix}
    1 & 0 \\
    0 & 1 \\
\end{pmatrix}
\label{eq::phi}
\end{equation}

and a {\bf process noise matrix} $Q$ that informs about the potential variation of the
state from epoch to epoch

\begin{equation}
\bm{Q} =
\begin{pmatrix}
    \sigma_x^2 & 0 \\
    0 & \sigma_y^2 \\
\end{pmatrix}
\label{eq::pnoise_matrix}
\end{equation}

The $\sigma_x$ and $\sigma_y$ values can be $0.2 \sim 0.5 m/s$ according to
the dynamics of the cart with the receiver (user equipment), as described in the dataset
documentation (\cite{ipin2022track8} and \cite{ipin2023track8}).

\end{itemize}

\begin{figure}[h]
\centering
\includegraphics[width=0.9\textwidth]{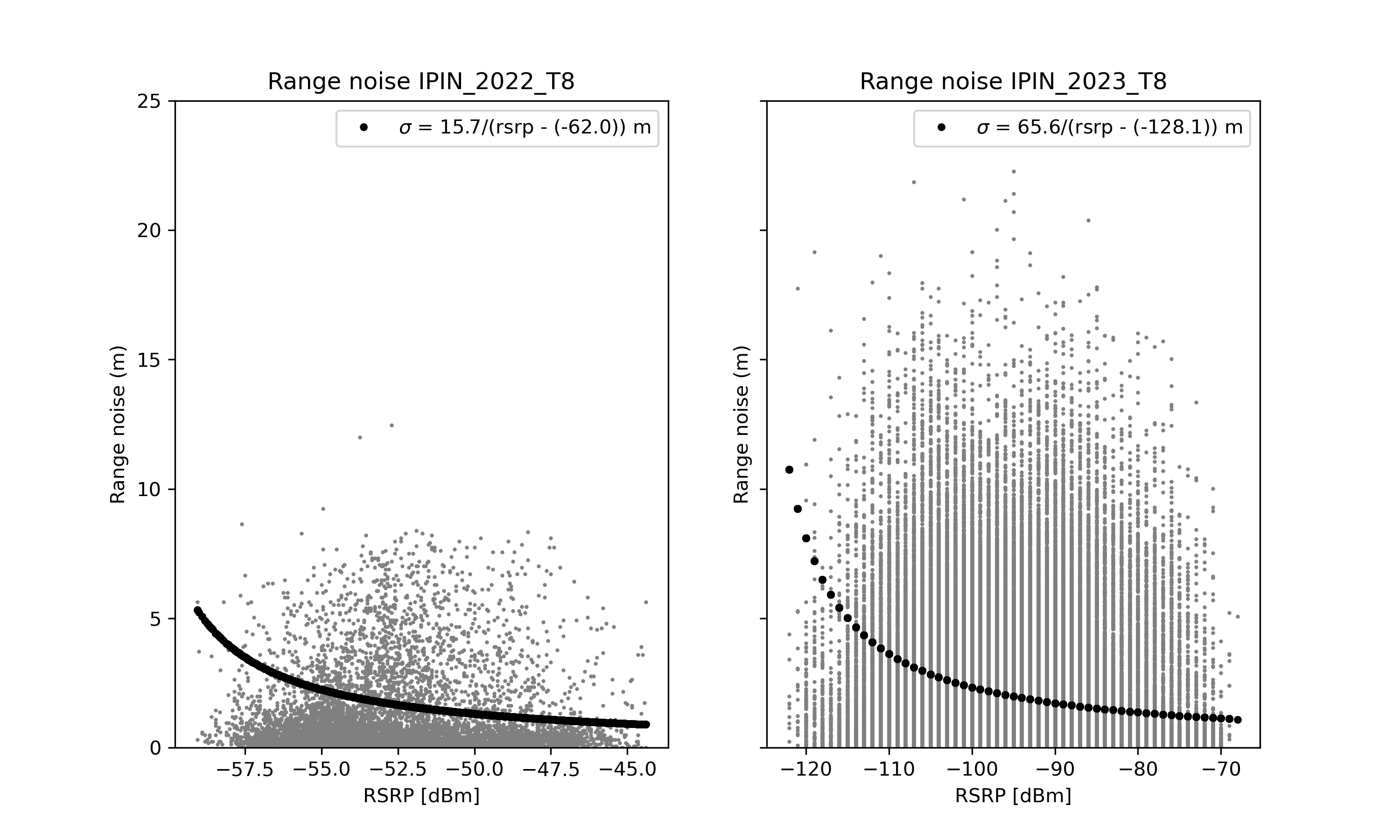}
\caption{Estimated range noises against the received power (RSRP). The figure
includes also an empirical measurement error model as black dots}
\label{fig::range_error_model}
\end{figure}

The measurement noise ($\sigma_{measurement}^2$) has been computed as a proportional
depedency on the inverse of the received power parameter (i.e. \texttt{RSRP},
in dBm), as used in previous works (e.g. \cite{huang2014tdoa}):

\begin{equation}
    \sigma_{measurement}(rsrp) = \frac{k}{rsrp-rsrp_0}
\label{eq::meas_error}
\end{equation}

where $k$ and $rsrp_0$ are parameters obtained from fitting the data points of
\texttt{RSRP} and range noise (see Figure \ref{fig::range_error_model}).
The range noise has been estimated by computing the detrended TOA measurements.
The detrending process removes the slow varying terms (i.e. geometry and biases),
while keeping the higher frequency terms such as the thermal noise
(the $\varepsilon_P$ term of Eq. \ref{eq::toa}). Albeit coarse, this model for
the measurement error based on the received power is enough to properly weight
the incoming measurements in the filter and avoid diverging estimations.

\section{Results and discussion}\label{sec::results}

This section contains the results obtained by processing the datasets prepared
for the two IPIN competition campaigns (2022 and 2023):

\begin{itemize}
\item IPIN 2022 5G positioning dataset: the dataset consists of measurements based
on 5G reference signals. A Huawei Mate 30 Pro smartphone send pulses to base
stations, which were responsible to compute the ToA and received power.
Full details available at \cite{ipin2022track8}.
\item IPIN 2023 5G positioning dataset: close setup to that of IPIN 2022 but with
additional nodes (base stations). Full details available at \cite{ipin2023track8}
\end{itemize}

The results contained in this section have been split into two parts: one
regarding the estimation of the $DTB$, and another one with the positioning
engine based on TDoA, that incorporates those $DTB$.

\subsection{DTB estimation}

This section presents the $DTB$ estimation results for all campaigns and datasets
available from the IPIN competition, employing the methodology outlined in Section
\ref{meas_model_and_dtb} and previously illustrated using data from the \texttt{D5}
session of the IPIN 2023 campaign.

Tables \ref{tab::dtb_ipin_2022} and \ref{tab::dtb_ipin_2023} display the $DTB$
values for IPIN 2022 and IPIN 2023, respectively. The IPIN 2023 campaign demonstrates
relatively homogeneous $DTB$ consistency across all nodes and datasets, with the
exception of Node 1, which exhibits a significantly different mean compared to other
nodes. Importantly, all mean values remain consistent across all sessions within
this campaign. This consistency is crucial for minimizing the frequency of $DTB$
calibration cycles.

In contrast, the previous IPIN 2022 campaign revealed substantial variations not
only between nodes but also between sessions, making it challenging to assign a
single $DTB$ per node for all sessions of the campaign (unlike the IPIN 2023 campaign).

Regardless, a clear observation is that the standard deviation of the $DTB$
($\sigma_{DTB}$) consistently remains around $2m$, further supporting the hypothesis
that the DTB can be considered a constant (at least within a measurement session).

\begin{table}[h]
\caption{DTB statistics (average and standard deviation) for IPIN 2022 dataset
(taking as reference node 1). Units are in meters.}\label{tab::dtb_ipin_2022}
\begin{tabular}{ccccc}
        &   \multicolumn{2}{c}{$\bar {DTB}$}  &    \multicolumn{2}{c}{$\sigma_{DTB}$}\\\cmidrule(r){2-3} \cmidrule(lr){4-5}
session &         D0 &        D1 &       D0 &       D1 \\
Node ID & & & & \\ \hline
0       & -20.9 & -0.8 & 2.3 & 1.5 \\
2       &  -5.3 & 25.8 & 1.0 & 1.1 \\
3       &  -4.6 & 22.0 & 2.1 & 1.2 \\ \hline
\end{tabular}
\end{table}

\begin{table}[h]

\caption{DTB statistics (average and standard deviation) for IPIN 2023 dataset
(taking as reference node 5). Units are in meters.}\label{tab::dtb_ipin_2023}
\begin{tabular}{ccccccccc}

        &     \multicolumn{4}{c}{$\bar {DTB}$}         &   \multicolumn{4}{c}{$\sigma_{DTB}$}\\\cmidrule(r){2-5} \cmidrule(lr){6-9}
session &         D2 &        D5 &        D6 &        D8 &       D2  &      D5 &       D6 &       D8 \\
Node ID & & & & & & & & \\ \hline
1       &  -6.5 & -7.7 & -6.6 & -7.4 & 1.9  &1.4 & 2.1 & 2.0 \\
2       &  18.7 & 17.6 & 17.7 & 18.0 & 1.8  &1.5 & 1.3 & 1.7 \\
3       &  18.8 & 18.8 & 18.6 & 18.7 & 1.6  &1.8 & 1.8 & 1.9 \\
4       &  17.4 & 17.4 & 17.0 & 17.5 & 1.8  &1.7 & 1.7 & 2.7 \\
6       &  21.0 & 20.3 & 20.7 & 20.8 & 2.5  &1.7 & 1.9 & 1.9 \\
7       &  20.5 & 20.9 & 21.3 & 20.8 & 2.0  &1.4 & 1.4 & 1.6 \\
8       &  20.3 & 21.0 & 20.8 & 20.9 & 2.1  &1.4 & 1.5 & 1.6 \\  \hline
\end{tabular}

\end{table}

\subsection{Positioning results}

\begin{figure}[h]
\centering
\includegraphics[width=0.7\textwidth]{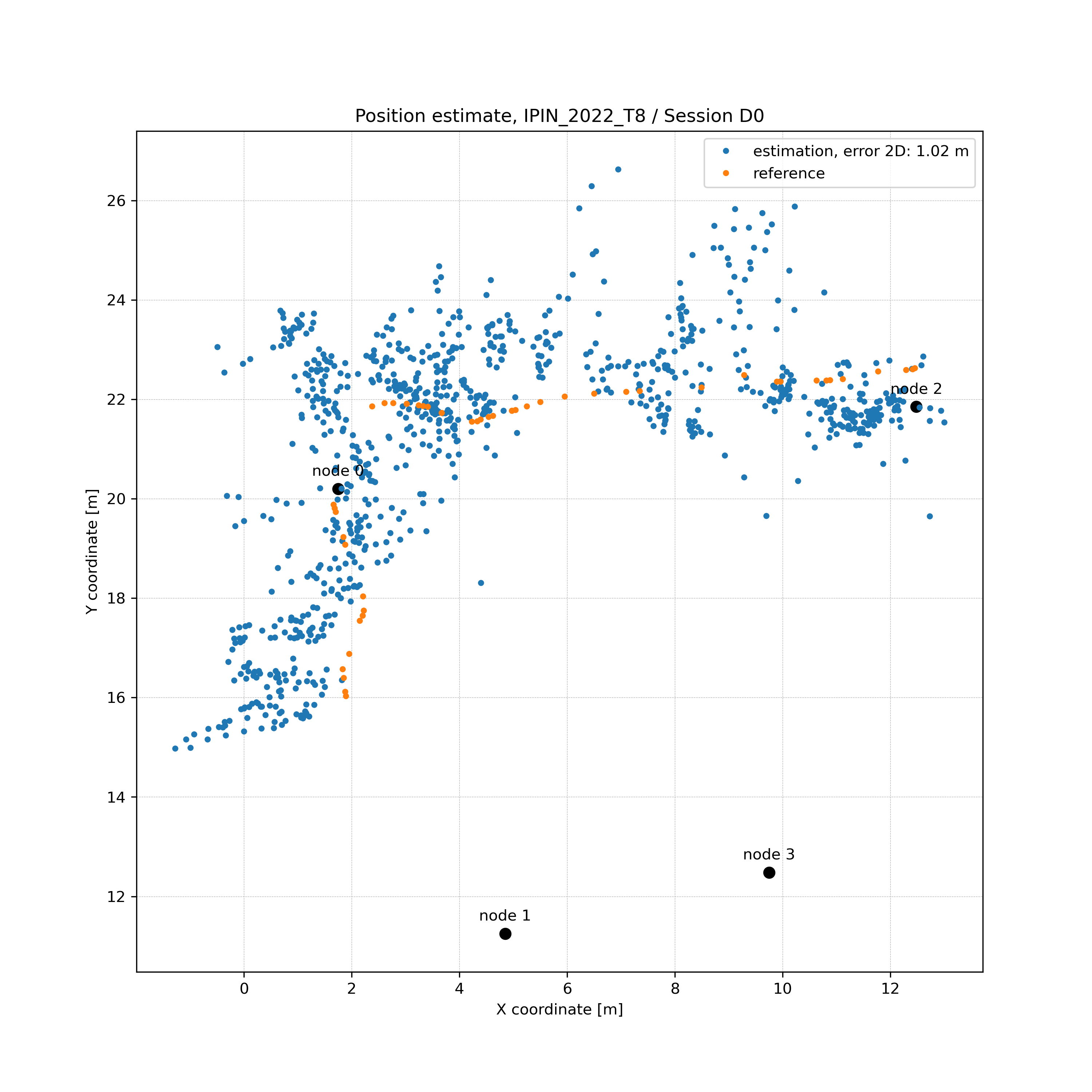}
\caption{Positioning results for the measurement session D0 of the IPIN 2022 campaign. Estimated and reference track are also shown in the plots.}
\label{fig::poserror_2022}
\end{figure}

\begin{figure}[h]
\centering
\includegraphics[width=0.7\textwidth]{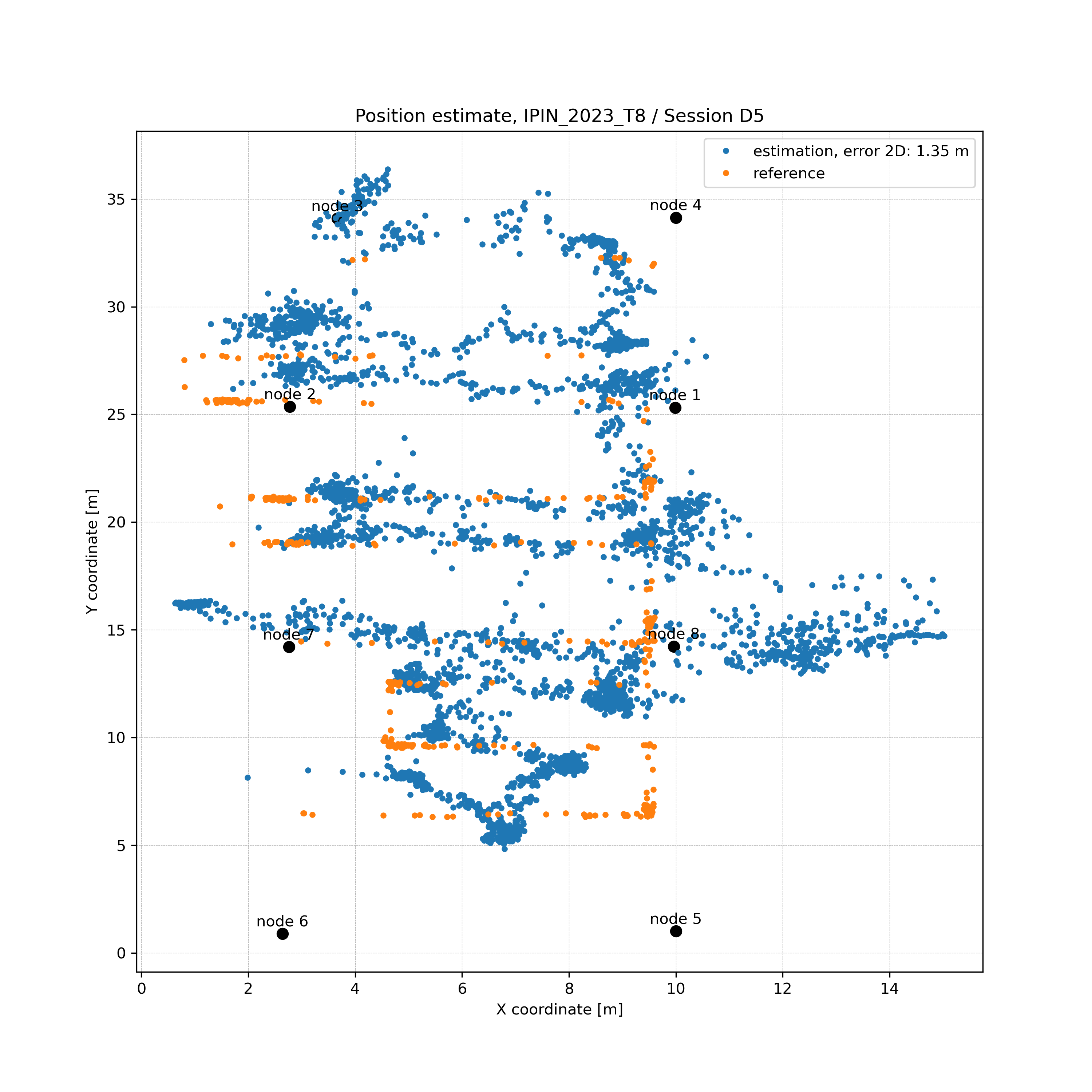}
\caption{Positioning results for the measurement session D5 of the IPIN 2023 campaign. Estimated and reference track are also shown in the plots.}
\label{fig::poserror_2023}
\end{figure}

Once the $DTB$ are computed, they can be used by the positioning engine implemented
via e.g. EKF, with the parameters outlined in Section \ref{sec::posengine}. The
positioning engine has been run for both the cases where the $DTB$ where considered
and for the case where they have not been considered. When $DTB$ are considered,
positioning errors of about 1.5 to 2 meters can be achieved (see Figures
\ref{fig::poserror_2022} and \ref{fig::poserror_2023}).
However, when no $DTB$ are handled (neither corrected nor estimated), the filter
diverges and no realistic estimates can be obtained. In this case, errors larger
than $100m$ where obtained and thus no statistics or figures are deemed relevant to
be included in this paper.

\begin{figure}[h]
\centering
\includegraphics[width=0.9\textwidth]{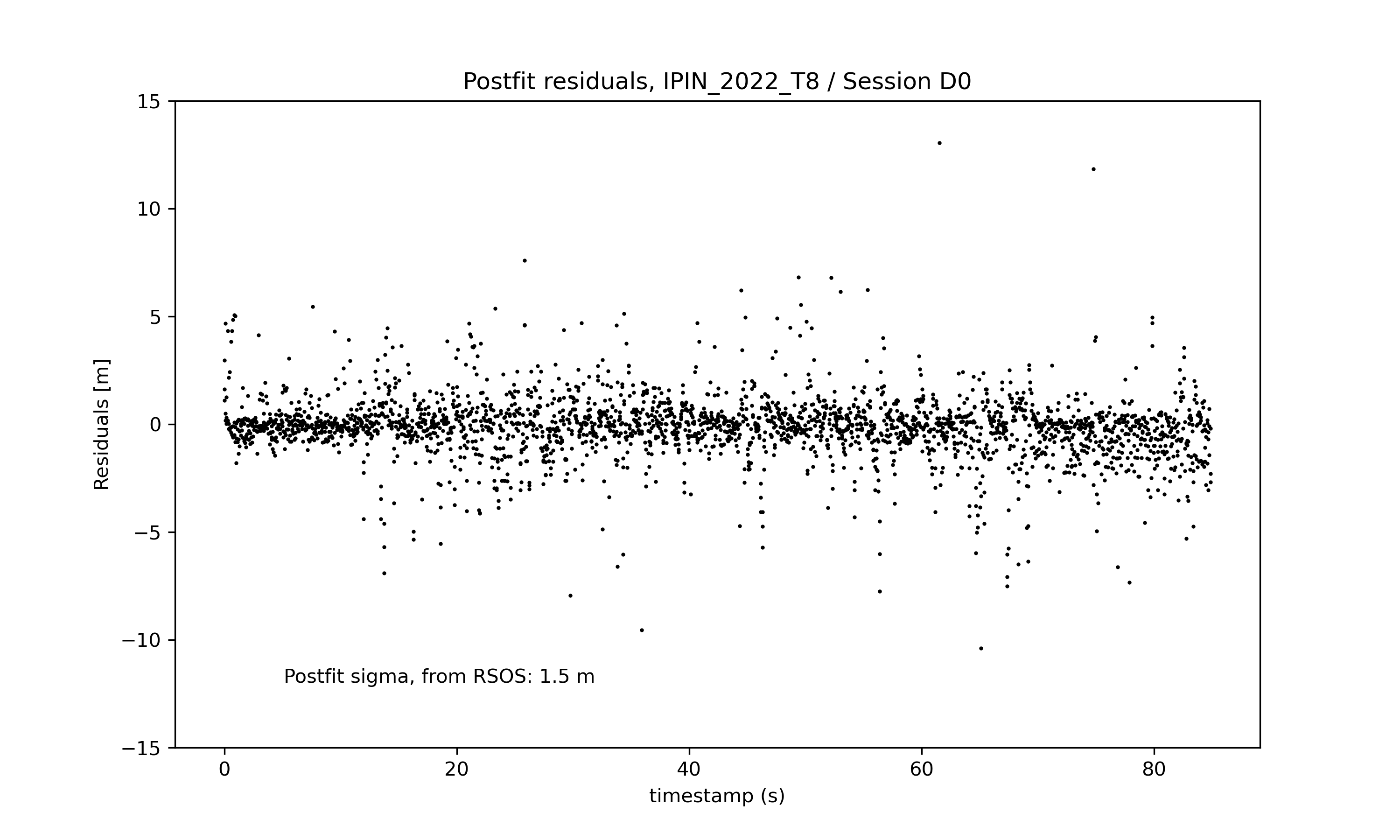} \\
\includegraphics[width=0.9\textwidth]{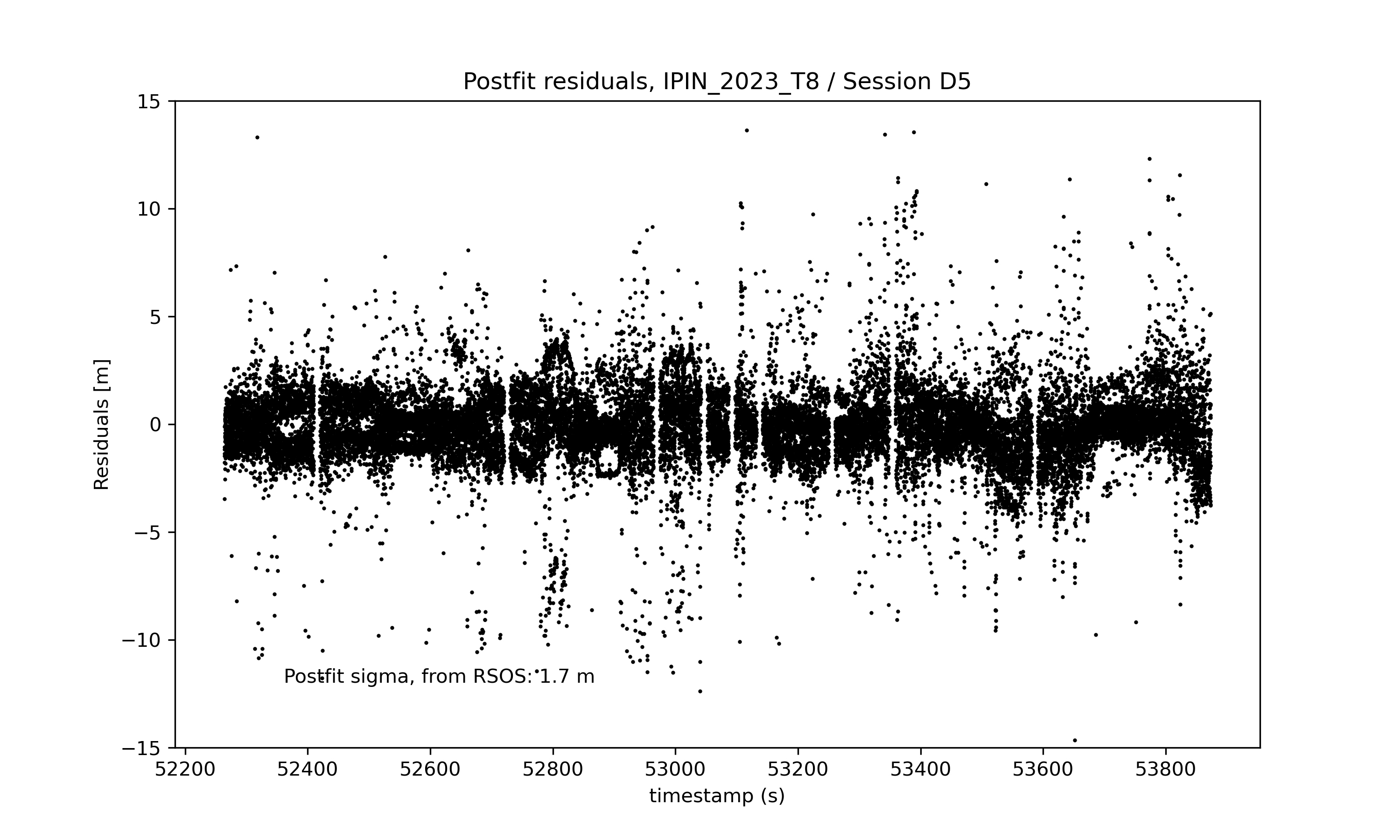} \\
\caption{Postfit residuals for the two selected sessions of the IPIN campaigns shown in Figures \ref{fig::poserror_2022} and \ref{fig::poserror_2023}. }
\label{fig::postfits}
\end{figure}

As an additional performance metric, the formal sigma and the variance computed
from the Residual Square Of Sum (RSOS) can be obtained. The former (i.e. $\sigma_{formal}$)
can be computed as the average of the trace of the postfit covariance
matrix. The RSOS-based sigma (i.e. $\sigma_{postfits}$) can be computed as:

\begin{equation}
\sigma_{postfits}^2 = \frac{1}{N - N_{param}}\sum_{i=1}^{N} e_i^2
\label{eq::sigma_postfits}
\end{equation}

where $e_i$ is the i-th postfit residual, $N$ is the number of points and $N_{param}$
is the number of parameters to be estimated (in this case the 2 position components).
The postfit residuals for the position solution of Figures \ref{fig::poserror_2022}
and \ref{fig::poserror_2023} are shown in Figure \ref{fig::postfits}. As expected,
these residuals show a Gaussian distribution centered at 0. The plots show also
some large values in the postfit residuals that could be removed for a potential
accuracy increase, but this feature has been left outside the scope of this work.

A comparison of the three performance metrics: \emph{true error} (obtained with
the comparison of the estimates with the reference trajectory provided in the dataset),
$\sigma_{formal}$ and  $\sigma_{postfits}$, is shown in Tables \ref{tab::err_ipin_2022}
and \ref{tab::err_ipin_2023} for both IPIN competitions. As a general rule, the
$\sigma_{formal}$ offers a rather too optimistic estimation of the error, while the one
based on the RSOS (i.e. postfits) is larger in all cases, thus providing an upper
boundary of the error.

\begin{table}[h!]
\caption{Positioning errors for IPIN 2022 dataset. Units are in meters.}\label{tab::err_ipin_2022}
\begin{tabular}{cccc}
    session  &  \emph{true error} & $\sigma_{formal}$ & $\sigma_{postfits}$ \\ \hline
    D0       &   1.0  &    0.9       & 1.5 \\
    D1       &   0.8  &    0.8       & 1.0 \\ \hline
\end{tabular}
\end{table}

\begin{table}[h!]
\caption{Positioning errors for IPIN 2023 dataset. Units are in meters.}\label{tab::err_ipin_2023}
\begin{tabular}{cccc}
    session  &  \emph{true error} & $\sigma_{formal}$ & $\sigma_{postfits}$ \\ \hline
    D2&  1.3   &   0.7   &     2.0\\
    D5&  1.3   &   0.7   &     1.7  \\
    D6&  1.3   &   0.7   &     1.8\\
    D8&  1.2   &   0.7   &     2.4\\  \hline
\end{tabular}
\end{table}

\section{Conclusion}\label{sec::conclusions}

This paper has addressed the critical issue of hardware biases in
wireless Time Difference of Arrival (TDoA) positioning. While TDoA offers
the advantage to remove the receiver
clock bias, the systematic bias due to the node hardware are still present
and should be dealt with. In particular, due to the differential nature of the TDoA
observations, the node hardware biases take the form of
\emph{Differential Transmitter Bias} ($DTB$) that can be estimated provided that
reference positions are available. In this case, the actual geometry can be
computed and subsequently removed from TDoA observations, thus leaving the
instantaneous $DTB$. These $DTB$ are in fact similar to the Differential
Code Biases ($DCB$) used when combining measurements in the GNSS use case.

Using real-world data from the IPIN conference competition, the paper shows that DTBs
exhibit significant stability, remaining essentially constant within a positioning
session, with an uncertainty on the order of 2 meters, comparable to the measurement
error. While these DTB values are likely hardware-dependent, they do not preclude the
potential for additional biases introduced by Non-Line-of-Sight (NLOS) conditions, a
particularly important consideration in indoor environments. The TDoA-based positioning
results included in the paper clearly highlight the necessity of incorporating
$DTB$ calibration to obtain meaningful position estimates. Adopting this strategy
can lead to positioning estimates with an accuracy of $1.5 \sim 2m$.

A potential limitation of this approach is that those $DTB$ need
to be provided to the user by the operator of the node network, similar to how
GNSS broadcasts satellites (i.e. nodes) orbits, clocks and biases
via the navigation message. However, the demonstrated stability of $DTB$s suggests
the possibility of estimating them within the positioning engine itself.
While this approach could eliminate the need for external $DTB$ provision,
it may increase the convergence time due to the increased number of unknowns
in the estimation filter (one parameter per node difference plus the
position coordinates). Future work could explore this on-the-fly estimation and
its impact in the convergence time of the solution.

\section{Data availability}\label{sec::data}

The data used in this work is publicly accessible through the IPIN conference competition
website (\url{https://competition.ipin-conference.org/}). Alternatively, the data has been
homogenized for the preparation of this paper and is available at the
\texttt{pnt\_datasets} Python package \cite{mgf2025pntdatasets}.

\section{Usage of artificial intelligence}\label{sec::ai}

Artificial intelligence (AI) tools were used to assist with stylistic polishing and
grammar correction in some sections of this paper (using the draft written by the author).
The author subsequently reviewed and edited all AI-generated text. The core
research, including ideation, code implementation, and analysis, was conducted
entirely by the author.

\bibliography{bibliography}

\end{document}